\documentclass[12pt]{article}
\usepackage{graphicx}

\begin{document}
\begin{titlepage}
\begin{center}
\bf{The Hubbard model:basic notions and selected applications}
\end{center}
\begin{center}
V.Celebonovic
\end{center}
\begin{center}
Institute of Physics,Pregrevica 118,11080 Zemun-Belgrade,Serbia
\end{center}

\begin{center}
e-mail$^{1}$: vladan@ipb.ac.rs

e-mail$^{2}$:vladan@fock.usc.edu    
\end{center}
 
J. of Optoelectronics and Advanced Materials,$\bf{11}$,pp.1135-1141(2009).

\begin{abstract}
	The aim of this paper is to present a self contained introduction to the Hubbard model and some of its applications. The paper consists of two parts: the first will introduce the basic notions of the Hubbard model,starting from the motivation for its development to the formulation of the Hamiltonian and some methods of calculation within the model. The second part will discuss some applications of the model to $1D$ and $2D$ systems,based on the combination of the author's results with those from the literature. 
\end{abstract}
\end{titlepage} 

\section{Introduction }
Physics and all other natural sciences have at least one point in common: they have to find a way to determine the characteristics of the objects they study. Generally speaking,this is done in various kinds of experiments. Performing an experiment means putting the systm under study in interaction with some kind of external probe,and then measuring the response of the system.If the interaction of the system under study with the external probe is weak, and if the particles constituting the system mutually interact weakly, the response of the system will be a linear function of the interaction strength. Such a situation is called the one electron picture in solid state physics. Many useful results have been obtained in solid state physics using this picture. 

What happens if the particles making up the system are correlated,or interact strongly with the external probe? Such materials were discovered in the last century;very well known examples are organic metals and high temperature superconductors. In materials like these,the one electron picture can not be applied and a new theoretical approach is needed. A typical shape of the temperature dependence of the resistivity of normal metals is presented in Fig.1,taken from [1].A similar curve for the organic conductors is represented in Fig.2 [2]. Even just a glance at these two figures shows that the temperature dependence of the resistivity for these two kinds of materials is drastically different,which proves that a new theoretical model is necessary for use in studies of the organic conductors.  The electrical resistivity of metals is modelled by the scattering of conduction electrons on phonons. Already early attempts to explain the resistivity of organic metals by the same physical process gave large discrepancies between the theoretical and experimental temperature dependencies of the resistivity of these materials [3].  

Apart the introduction,this paper has two further sections. The second one is devoted to the basic notions of the Hubbard model,while the third deals with some selected applications of this model. 
\newpage 
\begin{center}
\begin{figure}
\includegraphics[width=5cm,height=5cm]{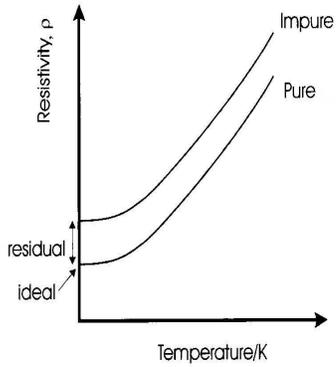}
\caption{\it{The temperature dependence of the resistivity of normal metals}}
\end{figure}
\end{center}

\bigskip
\begin{center}
\begin{figure}
\includegraphics[width=5cm,height=5cm]{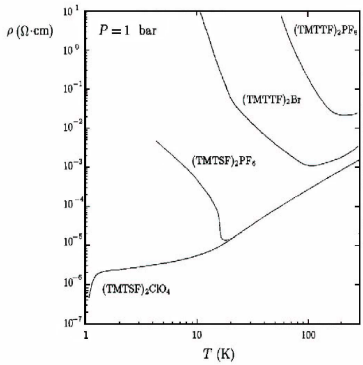}
\caption{\it{The temperature dependence of the resistivity of several organic metals}}
\end{figure}
\end{center}

\section{The Hubbard model}

\subsection{The basic notions}

Around the middle of the last century,one of the "`hot toppics"' in solid state physics was the phenomenon of the metal-insulator (MI) transition. Mott developed a theory of the MI transition,and showed what influence particle correlations can have on the results of the one electron picture [4]. Attempts to develop a microscopic model of the MI transition were important in the development of the Hubbard model (HM). 

The main "building block"  of the HM is a collection of atomic orbitals,and the main physical assumption is the idea of tight binding. Translated into common language,this means that the wave function of an electron is centered on the lattice site of an ion,and that any electron can "`hop"' one lattice spacing at a time. 

Denoting the orbitals by $\phi_{\mu}$ where $\mu\in[1,m]$,one can distinguish three kinds of them:

{\bf{Non-degenerate}} $\mu=1,2$  Such models are discussed in detail in [5],and they have only spin degenracy. The $s$ orbital is a typical example.

{\bf{Degenerate}} $\mu=1,2,..m$ and $m=2\times(2l+1)$. The symbol $l$ denotes the orbital angular momentum quantum number.Orbitals of this kind can be $p$,$d$,$f$...

{\bf{Multiple bands}} In this case several degenerate bands are combined.

The basic "`description"' of a model in statistical physics is its Hamiltonian. The Hamiltonian of the HM is given as the sum of two terms,the "free" kinetic term $H_{0}$ and the "`interaction"' term $H_{I}$

\begin{equation}
	H=H_{0}+H_{I}
\end{equation}
where
\begin{equation}
	H_{0}=\sum_{i,j}\sum_{\mu,\nu}T^{\mu,\nu}_{i,j}c_{i,\mu}^{+}c_{j,\mu}
\end{equation}

The hopping amplitude is given by
 
\begin{equation}
T^{\mu,\nu}_{i,j}=\int d^{3}x	\phi^{*}_{\mu}(x-R_{i})\times[-\frac{\hbar^{2}}{2m}\nabla^{2}+V(x)]
\end{equation}
In Eq.(3)$V(x)$ denotes the crystal ion potential felt by a single electron,while $\mu$ is an atomic orbital in an atom at lattice site $i$ and $\nu$ is an orbital in an atom at site $j$.The interaction term has a more complicated structure: 
\begin{equation}
	H_{I}=\frac{1}{2}\sum_{i,j,k,l}\sum_{\mu,\nu,\sigma,\tau}<i,\mu,j,\nu|\frac{1}{r}|k,\sigma,l,\tau>c_{i\mu}^{+}c_{j\nu}^{+}c_{k\sigma}c_{l\tau}
\end{equation}
Expressions of the form $<..||..>$ denote matrix elements of the Coulomb interaction between electrons on different lattice ions.

It can be expected that the most important role in the interaction term will be played by the electrons in orbitals on the same ion. In order to facilitate the applicability of Eq.(4) Hubbard introduced considerable simplifications.

He took into account only the matrix elements in which $i=j=k=l$ and assumed the existence of only one orbital. After these simpliications,the interaction term in the Hamiltonian takes the following form:
\begin{equation}
	H_{I}=U\sum_{i}n_{i,\uparrow}n_{i,\downarrow}
\end{equation}
with 
\begin{equation}
	U=<ii|\frac{1}{r}|ii>
\end{equation}
  The kinetic term can be expressed as:
  
\begin{equation}
	T_{ij}^{\mu\nu}=T_{0}\delta^{\mu\nu}\delta_{ij}+t^{\mu\nu}_{ij}
\end{equation}
The second term in Eq.(7) is non zero when the ions $i$ and $j$ are nearest neighbours. Assuming that $T_{0}=0$,which amounts to a change of gauge,and allowing for n.n hopping only,the Hamiltonian of a one-band Hubbard model becomes
\begin{equation}
	H=-t\sum_{<ij>,\sigma}(c_{i\sigma}^{+}c_{j}+c_{j\sigma}^{+}c_{i\sigma})+U\sum_{i}n_{i,\uparrow}n_{i,\downarrow}
\end{equation}
In one spatial dimension,the Hubbard Hamiltonian takes the following form 
\begin{equation}
H=-t\sum_{i,\sigma}(c_{i+1\sigma}^{+}c_{i\sigma}+c_{i\sigma}^{+}c_{i+1\sigma})+U\sum_{i}n_{i,\uparrow}n_{i,\downarrow}	
\end{equation}
Symbols of the form $c_{i\sigma}^{+}$ denote second quantisation operators creating an electron with spin $\sigma$ at a lattice site $i$,while $n_{i,\uparrow}$ is the number operator for the number of electrons at an ion on lattice site $i$ having spin up. 

The Hubbard model in $2D$ is mathematically much more complex. The Hamiltonian in this case has the form [6]: 
\begin{eqnarray}
H=-\sum_{ij}\sum_{\sigma}t_{ij}c_{i\sigma}^{+}c_{j\sigma}+\frac{1}{2}\sum_{ijkl}\sum_{\sigma\sigma'}<ij|v|kl>c_{i\sigma}^{+}c_{j\sigma'}^{+}c_{l\sigma'}c_{k\sigma}	
\end{eqnarray}
The Hubbard model may seem relatively simple,judging by the form of its Hamiltonian. However,it is only {\bf apparently} simple.Although nearly 60 years have passed since it was proposed, this model has been exactly solved only for the 1D case [7]. Results concerning low dimensional systems are very sensitive to details of the problem. For example,it was shown in [8] that the ground state of 1D systems is unmagnetized,assuming that the hopping is between nearest neighbours.Relaxing this assumption,and allowing for the next-nearest neighbour hopping,gives rise to ferromagnetic properties [9].

\subsection{Methods of calculation}
Immagine that the Hamiltonian H of a many-body system is at some sufficiently remote moment in the past perturbed by a time-dependent external field $h(t)$ . In real experiments,this external field can be high external pressure,which is at some moment turned on and increases with time.At time $t$ this Hamiltonian can be expressed as $H(t)=H+V(t)$,where $V(t)=h(t)A$,and the symbol $A$ denotes the parameter of the system with which the perturbing external field $h(t)$ is coupled.A common example of $A$ is the particle number density. 

As a consequence of the existence of the perturbation $h(t)$,the system under study is not "isolated" any more. This implies that the average value of the observable represented by the operator $A$ depends on the details of the perturbing field;solving such a problem is a complicated task. This problem can be reduced to the Linear Response Theory if the perturbation $h$ is small enough.

This idea is the foundation of the statistical mechanical theory of ireversible processes,proposed by Kubo [10]. The aim of this theory is to develop a scheme for the calculation of the kinetic coefficients for quantities such as the electrical and thermal conductivity. Kubo has shown that this calculation can be performed as a calculation of time correlation functions in equilibrium. From the viewpoint of pure theory,Kubo's theory solves the problem - it gives formal expressions for the required physical quantities. However,the expressions it gives are far too complex for application to real materials. 

Another method applicable to the calculation of the kietic coefficients is the so called "`memory function"' method,recently reviewed in [11]. This method is a logical continuation of the work by Kubo. It was practically developed in the 1970s,and within this method the electrical cnductivity can be calculated as follows:
      
\begin{eqnarray}
	\chi_{AB}(z)=<<A;B>>= -i\int_{0}^{\infty}\exp{izt}<[A(t),B(0)]> dt
\end{eqnarray} 

where $A=B=[j,H]$,$j$ denotes the current operator, and 

\begin{equation}
	\sigma(z) = i\frac{\omega_{P}^{2}}{4\pi z} [1-\frac{\chi_{z}}{\chi_{0}}]
\end{equation}

The symbol $\omega_{P}$ denotes the plasma frequency,which is given by $\omega_{P}^{2}=4\pi e^{2} n/m$,and where $e,n,m$ are the electron charge,number density and mass respectvely.  

\section{Selected applications}

\subsection{One-dimensional organic metals}

The general chemical formula of these materials is $(TMTSF)_{2}X$,where the formula  $(TMTSF)_{2}$ 
denotes a complicated chemical compound called di-tetra-methyl-tia-selena-fulvalene,and $X$ is any anion attached to it.A few examples of the anions which can be attached are $FSO_{3}$,$ClO_{4}$ or $NO_{3}$. After the name of the person who synthetized them for the first time,they became known as the Bechgaard salts. For a review of the field and some history,see [12]. 

Full details of the calculation of the electrical conductivity of the Bechgaard salts have been reviewed in [13]. The general conclusion drawn there is that the results of the calculation performed within the memory function method,using the Hubbard model and the Fermi distribution function are in good semi-quantitative agreement with experiments. This means that the general trend of the experimental data and some numerical values are successfully reproduced. 

Applications of the Fermi liquid theory to the Bechgaard salts are justified by the fact that these materials are not strictly one dimensional,but are in fact quasi-one-dimensional (Q1D). Recent experimental work has shown that at least some of these materials are quasi-two- dimensional [14].

The calculation of the electrical conductivity discussed in [14] had the disadvantage that the lattice constant was assumed to be equal to one. This detail helped to simplify the calculations, but at the same time excluded the possibility of taking into account the influence of variable external pressure on the conductivity.

The first step in the calculation [13] was to obtain an expression for the susceptibility $\chi$. This expression has the form of a sum. Performing this summation and then letting the immaginary part of the complex frequency tend to zero,it can be shown that the electrical conductivity is finally given by the following expression
\begin{equation}
\sigma_{R}(\omega_{0})=(1/2\chi_{0})(\omega_{P}^{2}/\pi)[\omega_{0}^{2}-(bt)^{2}]^{-1}(Ut/N^{2})^{2}\times S
\end{equation}

and the symbol $S$ denotes the following function

\begin{eqnarray}
S=42.49916\times(1+\exp(\beta(-\mu-2t)))^{-2}+78.2557\times
\nonumber\\
(1+\exp(\beta(-\mu+2t\cos(1+\pi))))^{-2}+(bt/(\omega_{0}+bt))\times
\nonumber\\	
(4.53316\times(1+\exp(\beta(-\mu-2t)))^{-2}+
\nonumber\\
24.6448(1+\exp(\beta(-\mu+2t\cos(1+\pi)))))^{-2})
\nonumber\\
\end{eqnarray}

The symbol $\mu$ denotes the chemical potential of the electron gas,

determined in [15] as:

\begin{equation}
\mu=\frac{(\beta t)^{6}(ns-1)\left|t\right|}{1.1029+.1694(\beta t)^{2}+.0654(\beta t)^{4}}	
\end{equation}
Obviously $lim_{n,s\rightarrow1}\mu$=0.     
In Eqs.(13)-(15),$\omega_{0}$ denotes the real part of the frequency,$\beta$ is the inverse temperature,$s$ the lattice constant and $n$ the band filling. The number of lattice sites is denoted by $N$. The static limit of the susceptibility is denoted by $\chi_{0}$. Equations (13) and (14) give the possibility for investigating the dependence of the conductivity on the temperature $T$,band filling $n$,frequency $\omega_{0}$ and hopping parameter $t$. However,these equations do not give the possibility for taking into account the influence of high external pressure on the conductivity.

As an example of the results obtained on the Bechgaard salts by application of Eqs.(13)-(15),the following figure gives the temperature dependence of normalized conductivity for two values of the band filling.
\newpage 
\begin{center}
\begin{figure}
\includegraphics[width=9cm,height=9cm]{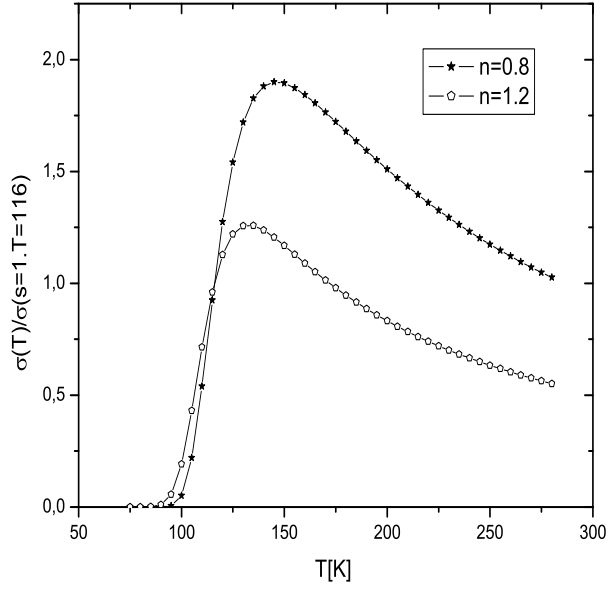}
\caption{\it{Normalzed conductivity of a Bechgaard salt for two values of the band filling}}
\end{figure}
\end{center}
The nonlinearity of the temperature dependence of the conductivity is clearly seen,as is the dependence on the band filling. This last conclusion has an interesting link to experiments. Namely,a band filling equal to 1 is usually taken to correspond to a pure specimen. Any deviation of $n$ from $1$ in fact means that the influence of doping on the conductivity is taken into account. 

\subsection{The conductivity of Bechgaard salts under pressure}

External pressure changes the value of the lattice constant of a material. Therefore,in order to take into account the influence of high pressure on the conductivity of the Bechgaard salts into account,the first step is to make a change of variables. All terms containing $k$ have to be replaced by terms containing $ks$. Summing the expression thus obtained within the first Brillouin zone,leads to a long impractical result for the real part of the susceptibility. It can be expressed in the form
\begin{equation}
	\chi_{R}(\omega)= \sum_{i}\frac{A_{i}}{\omega+q_{i} t} 
\end{equation}
where $q_{i}$ are numerical constants and the functions $A_{i}$ contain all the parameters of the problem except the frequency.The immaginary part of the dynamical susceptibility,denoted by $\chi_{I}$,is given by:
\begin{equation}
	\chi_{I}(\omega_{0})=-2(\frac{\omega_{0}}{\pi})P\int_{0}^{\infty}\frac{\chi_{R}(\omega)}{\omega^{2}-\omega_{0}^{2}} d\omega 
\end{equation}
where $P$ denotes the principal value of the integral.Performing the calculation it follows that
\begin{equation}
	\chi_{I}(\omega_{0})=\sum_{i}\frac{A_{i}}{\pi}\frac{\omega_{0}}{(q_{i}t)^{2}}\frac{1}{1-(\frac{\omega_{0}}{q_{i}t})^{2}}\times\ln(\frac{\omega_{0}}{q_{i}t})^{2} 
\end{equation}
under the assumption that $\omega_{0}^{2}>0$ and $(\omega_{0}/q_{i}t)^2<1$. As discussed in [15],the electrical conductivity is given by:
\begin{equation}
	\sigma_{R}=\frac{\omega_{P}^{2}\chi_{I}}{4\pi\omega_{0}\chi_{0}}
\end{equation}
The final result for the real part of the electrical conductivity of these materials has the following form:
\begin{equation}
	\sigma_{R}=\frac{1}{\chi_{0}}(\frac{\omega_{P}}{2\pi})^{2}\sum_{i}\frac{A_{i}}{(q_{i}t)^{2}}\times\frac{ln[\frac{\omega_{0}}{q_{i}t}]^{2}}{1-[\frac{\omega_{0}}{q_{i}t}]^{2}}
\end{equation}
Assuming that the derivative of the plasma frequency with pressure is small,it follows that
\begin{equation}
	\frac{\partial\sigma_{R}}{\partial s}=\frac{\omega_{P}^{2}}{4\pi\omega_{0}\chi_{0}}\times\frac{\partial\chi_{I}}{\partial s}
\end{equation}
Inserting Eq.(18) into Eq.(21) one gets the following general expression for the derivative of the electrical conductivity with respect to the lattice constant: 
 
\begin{equation}
\frac{\partial\sigma_{R}}{\partial s}=\frac{1}{\chi_{0}}(\frac{\omega_{P}}{2\pi})^{2}\sum_{i}\frac{1}{(q_{i}t)^{2}}\frac{ln[\frac{\omega_{0}}{q_{i}t}]^{2}}{1-[\frac{\omega_{0}}{q_{i}t}]^{2}}\times\frac{\partial A_{i}}{\partial s}	
\end{equation}
This is the general expression for the dependence of the electrical conductivity of the Bechgaard salts on the lattice constant.The sign of the derivative on the left-hand side of Eq.(22) obviously depends on the sign of $\partial A/\partial s$. The number of terms which can be taken into account in any real applications of Eqs.(20) and (22) is limited mainly by the avaliable computing power.

For the purpose of illustrating the applicability of Eq.(20),the sum in it was limited to the first $9$ terms. The conductivity was calculated for the followingvalues of the parameters: $N=150$,$t=0.015$,$U=4.5 t$,$\omega_{0}=0.04$. The conductivity was calculated for different values of the lattice constant $s$.For every value of the lattice constant,the temperature at which the maximal value of the conductivity occurs,and the value of the conductivity itself were noted.The conductivity was normalized to unity at the point $s=1$,$T=116K$. The resulting plots are shown on  figures 4 and 5. On both figures,the compression is defined by the relation $\Delta s/s_{0}=(s_{0}-s)/s_{0}$. The starting,arbitrarily chosen value of the lattice constant is denoted by $s_{0}$. Both figures have been prepared for the same values of the band filling: $n=0.8$ and $n=1.2$.    
\begin{center}
\begin{figure}
\includegraphics[width=8cm,height=8cm]{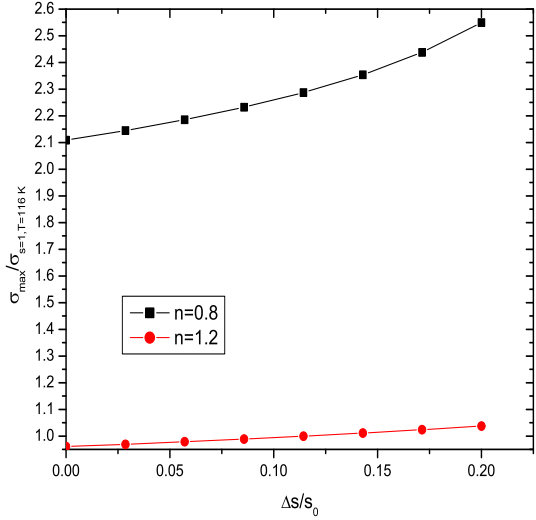}
\caption{\it{The maximal conductivity as a function of compression for two values of the band filling}}
\end{figure}
\end{center} 

\begin{center}
\begin{figure}
\includegraphics[width=8cm,height=8cm]{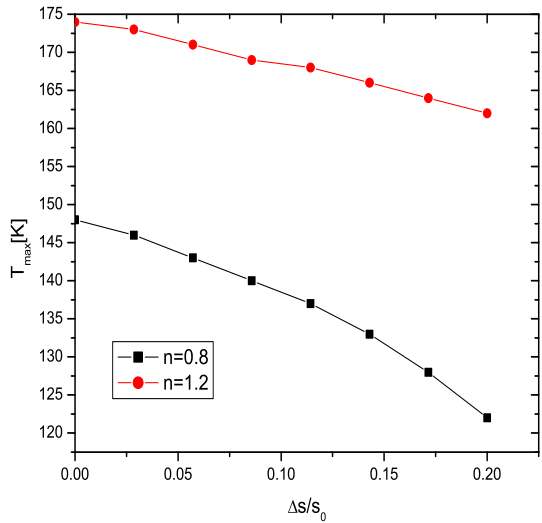}
\caption{\it{The temperature of maximal conductivity as a function of compression for two values of the band filling}}
\end{figure}
\end{center}

Clearly,for both chosen values of the band filling, the electrical conducctivity increasess with increasing compression. The increase is steeper for a band filling smaller than $1$. The temperature at which the conductivity attains its maximum for a given value of the compression diminishes with increasing compression. The decrease is again steeper for $n=0.8$. 

The electrical conductivity of the Bechgaard salts has been discussed here within the one-dimensional Hubbard model. The behaviour of these materials is a result of the influence of two competing factors which contribute to the Hamiltonian of the model: intersite hopping of the electrons and their localisation on the lattice nodes. Both of these factors are pressure dependent. The hopping integral $t$ is pressure dependent,as its definition contains the overlap of the electronic wave functions  on two adjacent sites,and the mutual distance of ajacent sites shrinks under increasing pressure. An electron localised on a lattice site can be though of as a particle bound in a finite potential well. It is known from various studies that such systems tend to get excited and finally ionised when exposed to sufficiently high external pressure (for example [16],[17] and references therein). This result is a theoretical "`corner stone" of an experimental method for measuring high static pressure [18]. 

\subsection{Two dimensional systems}
The Hubbard model is most often applied to two dimensional systems by using the so called Determinant Quantum Monte Carlo Method (DQMCM) developed in the 1980s at the University of California at Santa Barbara.A breif review of the method is presented in [19] and references given there.The calculation of the partition function is also discussed therein.Figure 6 shows one of the interesting results  obtained by applying the DQMCM to the 2D Hubbard model-the conductivity is plotted as a function of temperature $T$ for various values of the disorder $\Delta_{t}$. 
\begin{center}
\begin{figure}
\includegraphics[width=8cm,height=8cm]{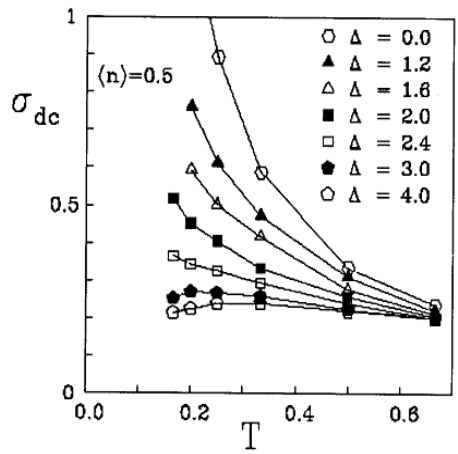}
\caption{\it{Temperature dependence of the conductivity $\sigma_{DC}$ for various values of the disorder $\Delta$ at $U=4$ for $<n>=0.5$.Calculations were performed on an $8x8$ lattice,and data points are averages over 4 realizations of a given disorder.}}
\end{figure}
\end{center}

Figure 6 shows the temperature dependence of the conductivity of a disordered 2D Hubbard model for variou svalues of the parameter $\Delta_{t}$ characterizing the disorder. Visibly,the shape of the curve changes for values of $\Delta_{t}$ between 2.4 and 3. This change is interpreted as a sign of the metal to insulator transition driven by disorder.As the conductivity is related to optical properties of a material,a similar kind of behaviour can be expected for reflectivity.An interes ting  conclusion has been reached on the behaviour of disordered two dimensional systems in a magnetic field. It has been shown [21] that a Zeeman magnetic field reduces the conductivity of a conducting disordered 2D system,under the assumption that the disorder strength is fixed and that the field is varied. AFter some value of the temperature,the field becomes temperature independent. This conclusion was reache dtheoretically an donly a posteriori related to already existing experimental data [22]. To make the result seven more complex,it turned out in [22] that the value of the resistivity at the metal-insulator transition was density dependent.This automatically means that it can be "`tuned"' by the application of high external pressure.

Before the end of this paper,a few words are in order on its subject. Namely,so far the word "membrane" has not appeared in i,so at first sight it might it may seem to have "missed the subject".The answer is that membranes (biological or artificial) are always two dimensional,so everything stated in this paper concerning 2D electronic systems should (in principle at least) be applicable to membranes. Similarily,one dimensional systems can be viewed as apecial membranes,which have their length much bigger than their width. Accordingly,the mathematical and physical considerations presented in this paper are,perhaps with some small modifications,applicable to problems with membranes.

\section{Conclusions} 

This paper was prepared with two precise aims: to present an as much as possible self contained introduction to the basic notions cpncrning the Hubbard model, and to discuss to a limited extent its selected applications. In the part concerning the Hubbard model itself,motives for its development are explained are expressions are given for the Hamiltonian in the $1D$ and $2D$ cases. Concerning the electrical conductivity,explicite expressions are given for it,within a particular theoretical approach. Accordingly,the interested reader can start a calculation on his/her own. The question of the thermal conductivity of the Hubbard model has been deliberately left out,as it will be the subject of further work. The applications discussed reffer to $1D$ and $2D$ electronic systems;as particular examples,we have considered $Q1D$ organic metals and $2D$ electronic systems. 

In the $1D$ case,this paper contains results on the electrical conductivity of the Bechgaard salts,at first without and then with taking the influence of high external pressure on their conductivity. BOth $1D$ and $2D$ systems can be considered as special cases of membranes. 

In closing,note that although the Hubbard model was proposed in the middle of the last century, and solved for the $1D$ case back in 1968.,it still offers numerous possibilities for active research work. It is hoped that this paper will contribute to the spread of interest in this model and its applicability.

\section{Note} 
This is the text of an invited lecture presented by the author at the joint Bulgarian-British School in solid state physics,held near Varna (Bulgaria), August 31.,-September 5.,2008.


\section{Acknowledgement} 

The preparation of this work was financed by the Ministry of Science and Technology of Serbia under its project 141007. 
  

{}

\end{document}